\documentclass[aps,prl,superscriptaddress,twocolumn,longbibliography]{revtex4-1}
\usepackage{bbm}
\usepackage{graphicx}
\usepackage{dcolumn}
\usepackage{bm}
\usepackage{subfigure}
\usepackage{amsmath}
\usepackage{feynmf}
\usepackage{hyperref}

\usepackage{attachfile}

\newcommand{\bk}{\boldsymbol k}

\newcommand{\zhongbo}{\color {black}}
\newcommand{\zb}{\color {black}}
\usepackage{times}

\begin{document}
\title{Anomalous Linear and Quadratic Nodeless Surface Dirac Cones \\in Three-Dimensional Dirac Semimetals}

\title{Anomalous Linear and Quadratic Nodeless Surface Dirac Cones \\in Three-Dimensional Dirac Semimetals}

\author{Dongling Liu}
\altaffiliation{These authors contributed equally to this work.}
\affiliation{Guangdong Provincial Key Laboratory of Magnetoelectric Physics and Devices,
State Key Laboratory of Optoelectronic Materials and Technologies,
and School of Physics, Sun Yat-sen University, Guangzhou 510275, China}

\author{Xiao-Jiao Wang}
\altaffiliation{These authors contributed equally to this work.}
\affiliation{Guangdong Provincial Key Laboratory of Magnetoelectric Physics and Devices,
State Key Laboratory of Optoelectronic Materials and Technologies,
and School of Physics, Sun Yat-sen University, Guangzhou 510275, China}

\author{Yijie Mo}
\affiliation{Guangdong Provincial Key Laboratory of Magnetoelectric Physics and Devices,
State Key Laboratory of Optoelectronic Materials and Technologies,
and School of Physics, Sun Yat-sen University, Guangzhou 510275, China}

\author{Zhongbo Yan}
\email{yanzhb5@mail.sysu.edu.cn}
\affiliation{Guangdong Provincial Key Laboratory of Magnetoelectric Physics and Devices,
State Key Laboratory of Optoelectronic Materials and Technologies,
and School of Physics, Sun Yat-sen University, Guangzhou 510275, China}

\date{\today}

\begin{abstract}
Surface Dirac cones in three-dimensional topological insulators have generated tremendous and enduring interest
for almost two decades owing to hosting a multitude of exotic properties.
In this work, we unveil the existence of two types of anomalous surface Dirac cones in
three-dimensional Dirac semimetals.
These surface Dirac cones {\zb are located at the surfaces perpendicular to the
rotation symmetry axis}, and are found to display a number of features remarkably different from that
in topological insulators. The most prominent one is the absence of singular Dirac node.
In addition, the spin textures of these nodeless surface Dirac cones are
found to exhibit a unique two-phase-angle dependence, leading to the presence of two different
winding numbers in the orbital-resolved spin textures, which is rather different from
the well-known spin-momentum locking in topological insulators. Despite the absence
of Dirac node, we find that the two types of surface Dirac cones are also characterized by quantized $\pi$
Berry phases, even though one of them takes a quadratic dispersion.
In the presence of time-reversal-symmetry-breaking fields, we find that
the responses of the surface and bulk Dirac cones display an interesting bulk-surface
correspondence. The uncovering of these nodeless surface Dirac cones broadens our understanding of
the topological surface states and bulk-boundary correspondence in
Dirac semimetals, and also lays down the basis for studying unconventional Dirac physics.
\end{abstract}

\maketitle

Since the rise of graphene and topological insulators (TIs),
the exploration of  Dirac-cone band structures has continued to be at
the frontier of a number of disciplines~\cite{Novoselov2004,Kane2005a,Kane2005b,Bernevig2006a,Bernevig2006c,hasan2010,qi2011,Vafek2014,Armitage2018,lu2014topological,Ozawa2019review,Ma2019,Xue2022}.
The great interest in
Dirac-cone band structures lies in many aspects, such as their relativistic
linear dispersions~\cite{Novoselov2005massless}, their fundamental connection with topology~\cite{jackiw1976b,Ryu2010,Chiu2015RMP}, and being the sources
of a diversity of unconventional responses~\cite{cheng2010,Tse2010,Garate2010,Xiong2015anomaly,Wu2016TI,Li2016CME,Yuan2018,Yuan2020}.
The Dirac cones can be roughly classified into two classes,
gapped or gapless, with the former (latter) effectively described
by a massive (massless) Dirac Hamiltonian~\cite{shen2013topological,
bernevig2013topological}.
A fundamental difference between them is that the gapless Dirac cones
carry a symmetry-protected band degeneracy (known as Dirac node or point) that acts as a topological charge.
The discovery of an odd number of 2D gapless Dirac cone on the surface
of a 3D strong TI~\cite{Xia2009bise,Hsieh2009bise,Chen2009bise,Sato2010Dirac,Chen2010TI}
has attracted particular interest since it not only provides an exception to the
fermion-doubling problem~\cite{Nielsen1981nogo,Nielsen1981a,Nielsen1981b},
but also realizes a class of  unconventional
metals with many intriguing properties. {\zhongbo Notable properties associated with a gapless surface Dirac cone (SDC)
include the quantized $\pi$ Berry phase
that can lead to weak antilocalization in transport~\cite{He2011localization,Lu2011antilocalization},
and the spin-momentum-locking Fermi surface~\cite{Hsieh2009helical,Souma2011texture}
that can create non-Abliean Majorana zero modes when
superconductivity is brought in~\cite{fu2008,Sun2016Majorana,wang2018evidence}. Moreover,
when the gapless SDC is gapped by certain time-reversal symmetry (TRS) breaking field,
half-integer quantum Hall effects
as well as topological electromagnetic effects can be observed~\cite{Qi2008TFT,Essin2009,Mong2010,Mogi2022}}.

TIs build a common picture through the bulk-boundary correspondence that
the 2D gapless SDCs are decedent from the 3D gapped Dirac cones in the bulk~\cite{Zhang2009bise,Liu2010model}.
However, this does not mean that gapless SDCs can only appear in TIs.
As an intermediate phase between TIs and normal insulators,
3D Dirac semimetals (DSMs) with band-inverted structure and rotation symmetry in fact can also support
an odd number of 2D gapless Dirac cones on a given surface. {\zhongbo This fact was first noticed when Kargarian {\it et al.}
revealed that the Fermi arcs in DSMs could deform into Fermi loops
~\cite{Kargarian2016}, which implies the possibility of the existence of SDCs in DSMs.}
Later Yan {\it et al.} analytically derived the low-energy
Hamiltonian describing the surface states and
showed how the gapless SDCs arise~\cite{Yan2020vortex}.
All these studies, however, are restricted to the side surfaces parallel
to the rotation axis where the bulk Dirac nodes are located,
owing to the primary interest in Fermi arcs and the fact that Fermi arcs only exist
on the surfaces where the projections of the bulk
Dirac nodes do not overlap~\cite{xu2015observation}.

\begin{figure}[t]
\centering
\includegraphics[width=0.4\textwidth]{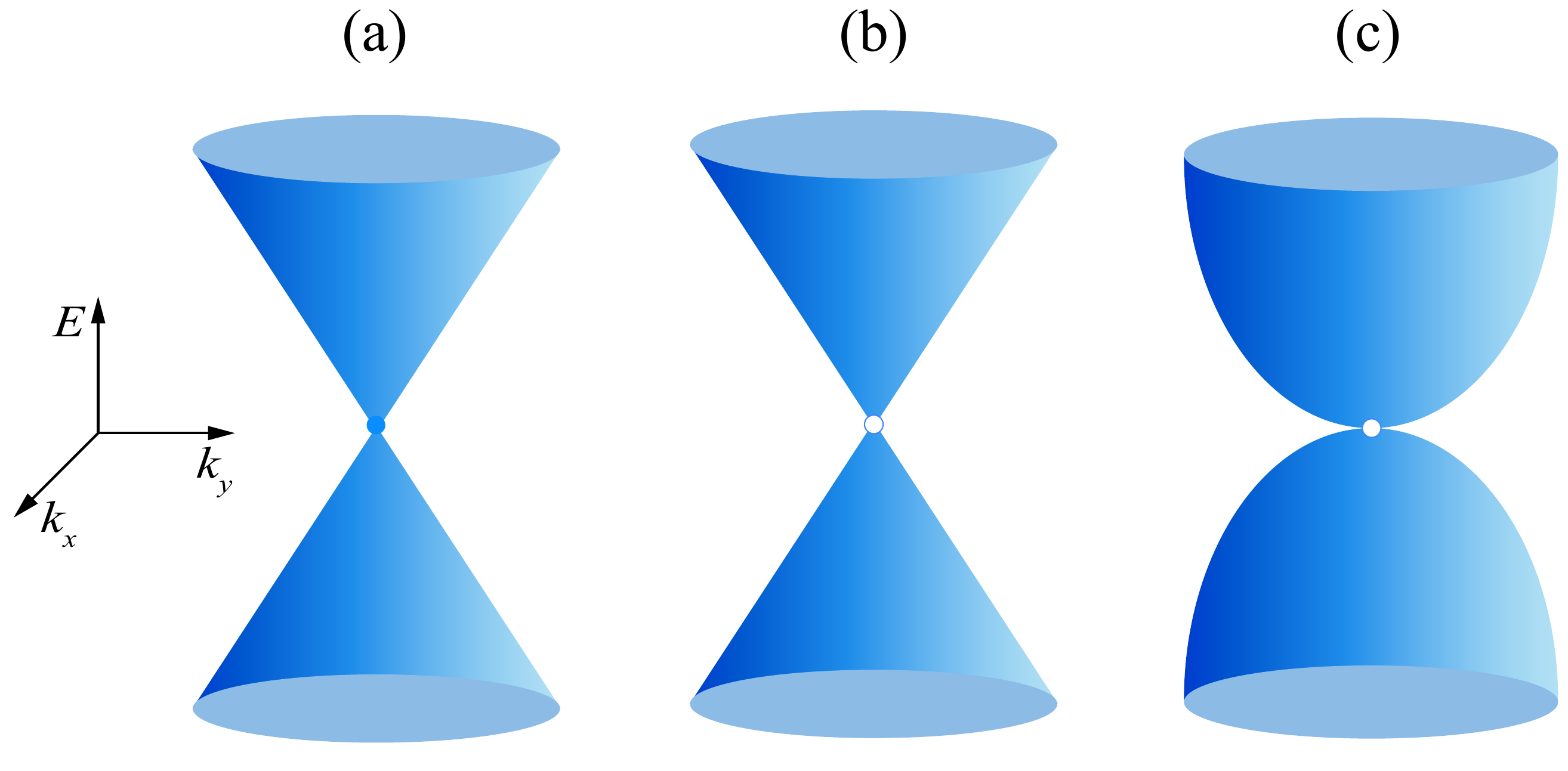}
\caption{(color online) Schematic diagrams of three types of gapless SDCs. (a)
Linear SDC in TIs, the Dirac node (a Kramers degeneracy) denoted by a solid dot is enforced by
TRS. (b) and (c) are respectively the linear and quadratic nodelss SDCs in the opposite-parity and
same-parity DSMs.   The open circles
in (b) and (c) represent the absence of Dirac node.  }
\label{sketch}
\end{figure}

Recently, a remarkable experiment
reported the observation of 2D gapless SDCs
in some iron-based superconducting compounds with 3D bulk Dirac nodes protected by
$C_{4z}$ rotation symmetry~\cite{zhang2019multiple}. Notably, the 2D gapless Dirac
cones are located on the surface where the projections of the
bulk Dirac nodes overlap, revealing that the largely-overlooked top and bottom surfaces
perpendicular to the rotation axis also carry interesting topological surface states in DSMs.
Inspired by this experiment, we consider two representative types of 3D DSMs protected
by $C_{4z}$ rotation symmetry and explore the topological surface states
on the top and bottom surfaces. Remarkably, we find that the gapless Dirac cones found
on these surfaces display a number of features sharply distinct from the SDCs
in TIs. The most evident difference is the absence of Dirac node in them,
as illustrated in Fig.\ref{sketch}. The spin textures of these nodeless SDCs are
found to have a unique two-phase-angle dependence enforced by a subchiral symmetry, rather different from
the one-phase-angle dependence exhibited in TIs. Furthermore, despite the absence
of Dirac node, we find that the two types of SDCs are also characterized by quantized $\pi$
Berry phases, even though one of them has a quadratic dispersion.
In the presence of TRS-breaking fields, we find that
the responses of the surface and bulk Dirac cones display an interesting
bulk-surface correspondence.

{\it Linear nodeless SDCs in the opposite-parity DSM.---} DSMs
are materials whose conduction and valence bands cross at some isolated points (Dirac nodes)
in the Brillouin zone~\cite{young2012dirac,wang2012e,wang2013three,liu2014discovery,Liu2014b,Borisenko2014,neupane2014observation}. Depending on whether the band crossings occur between
bands with opposite parity or same parity, DSMs can be roughly
divided into two classes~\cite{Qin2019vortex}. For the convenience of discussion, we dub the class involving bands
with opposite (same) parity as opposite-parity (same-parity) DSMs.

Let us first consider the opposite-parity DSM. Focusing on a cubic-lattice realization,
the minimal model  is given by~\cite{wang2013three,Yang2014}
\begin{eqnarray}
\mathcal{H}(\bk)&=&(m-t\cos k_{x}-t\cos k_{y}-t_{z}\cos k_{z})\sigma_{z}s_{0}\nonumber\\
&&+\lambda (\sin k_{x}\sigma_{x}s_{z}-\sin k_{y}\sigma_{y}s_{0})\nonumber\\
&&+\eta_{1}(\bk_{s})\sin k_{z}\sigma_{x}s_{x}+\eta_{2}(\bk_{s})\sin k_{z}\sigma_{x}s_{y},\label{type1}
\end{eqnarray}
where $\bk_{s}=(k_{x},k_{y})$ denotes the $xy$-plane momentum,
$\eta_{1}(\bk_{s})=\eta_{1}(\cos k_{x}-\cos k_{y})$, $\eta_{2}(\bk_{s})=\eta_{2}\sin k_{x}\sin k_{y}$,
$\sigma_{i}$ and $s_{i}$ are Pauli matrices in orbital and spin space, and $\sigma_{0}$ and $s_{0}$ are
the corresponding identity matrices. For notational simplicity, the lattice constants are set to unity
throughout. Without loss of generality, below we consider all parameters in Eq.(\ref{type1}) to be positive, and $|m-2t|<t_{z}<m$. Accordingly,
a band inversion occurs at the time-reversal invariant momentum
$\boldsymbol{\Gamma}=(0,0,0)$, and there are  two Dirac nodes located at $\bk_{D,\pm}=\pm(0,0,k_{D})$
with $k_{D}=\arccos (m-2t)/t_{z}$. It is noteworthy that the existence and the locations
of the bulk Dirac nodes do not depend on the two $\eta$ terms. However, as we shall show below,
the $\eta$ terms have rather remarkable effects on the topological surface states.

When $\eta_{1}$ and $\eta_{2}$ vanish,
the Hamiltonian (\ref{type1}) at a given $k_{z}$ is characterized by a $Z_{2}$ invariant~\cite{Bernevig2006c},
and describes a 2D TI for $|k_{z}|<|\bk_{D}|$, and a normal insulator
for $|k_{z}|>|\bk_{D}|$. For this situation, the DSM can  be
regarded as a stacking of 2D TIs in the $z$ direction.
Accordingly, the surface states only exist on the side surfaces, and
the iso-energy contours of these surface states form the so-called
Fermi arcs.  Once $\eta_{1}$ and $\eta_{2}$ become finite, the dispersions of the surface
states on the side surfaces change dramatically~\cite{Kargarian2016,Yan2020vortex,Kargarian2018,Le2018coexist,Qin2023DSM}, leading to the change
of the Fermi-arc connectivity and the arising
of SDCs that can have nontrivial interplay with superconductivity~\cite{Kobayashi2015,Majid2022surface,Wu2022nodal}.
Furthermore, it has been recognized that the $\eta$ terms can also give arise to
gapless hinge states~\cite{Szabo2020}, a hallmark of second-order topology. These findings
have one after another deepened our understanding on the bulk-boundary correspondence
of DSMs. Now we show that our understanding remains incomplete.

To intuitively show that 2D gapless Dirac cones also exist on the top and bottom surfaces,
we first introduce a set of momentum-dependent Pauli matrices, namely,
\begin{eqnarray}
\tilde{s}_{x}&=&\cos\theta_{\bk_{s}}s_{x}+\sin\theta_{\bk_{s}}s_{y},\nonumber\\
\tilde{s}_{y}&=&-\sin\theta_{\bk_{s}}s_{x}+\cos\theta_{\bk_{s}}s_{y},\nonumber\\
\tilde{s}_{z}&=&s_{z},
\end{eqnarray}
where $\theta_{\bk_{s}}=\arg[\eta_{1}(\bk_{s})+i\eta_{2}(\bk_{s})]$. In this
work, two phase angles will be involved, one is $\theta_{\bk_{s}}$, and
the other is $\phi_{\bk_{s}}=\arg(\sin k_{x}+i\sin k_{y})$.
When considering the continuum counterpart of the lattice Hamiltonian,
these two phase angles are implicitly assumed to take
the corresponding continuum forms (e.g., $\phi_{\bk_{s}}=\arg(k_{x}+i k_{y})$).

It is easy to verify that this set of Pauli matrices also
satisfies $[\tilde{s}_{i},\tilde{s}_{j}]=2i\epsilon_{ijk}\tilde{s}_{k}$ and
$\{\tilde{s}_{i},\tilde{s}_{j}\}=2\delta_{ij}s_{0}$ for $ i,j\in\{x,y,z\}$.
Using them, the Hamiltonian can be rewritten as
\begin{eqnarray}
\mathcal{H}(\bk)&=&(m-t\cos k_{x}-t\cos k_{y}-t_{z}\cos k_{z})\sigma_{z}s_{0}\nonumber\\
&&+\lambda (\sin k_{x}\sigma_{x}\tilde{s}_{z}-\sin k_{y}\sigma_{y}s_{0})\nonumber\\
&&+\eta(\bk_{s})\sin k_{z}\sigma_{x}\tilde{s}_{x},
\end{eqnarray}
where $\eta(\bk_{s})=\sqrt{\eta_{1}^{2}(\bk_{s})+\eta_{2}^{2}(\bk_{s})}$. The above form resembles
the minimal model for 3D TIs~\cite{Zhang2009bise}, suggesting
the existence of gapless Dirac cones on the $z$-normal surfaces if $\eta(\bk_{s})$ is nonzero.
In this form, it is also easy to see that there exists a unitary operator anticommuting with the
Hamiltonian, i.e., $\{\mathcal{C},\mathcal{H}\}=0$, with $\mathcal{C}=\sigma_{x}\tilde{s}_{y}$.
Conventionally, such an anticommutation relation suggests that the Hamiltonian has chiral symmetry. However,
here is not the case, simply because the operator $\mathcal{C}$ is not a constant operator but
depends on partial components of the momentum vector. Such an algebraic property
was recently discussed and dubbed subchiral symmetry in Ref.\cite{Mo2023}. An
important conclusion from Ref.\cite{Mo2023} is that the subchiral symmetry operator
itself admits topological characterization, and  its topological property will impart
into the spin texture of the topological boundary states. Apparently, here
$\tilde{s}_{y}$ displays a nontrivial winding as $\bk_{s}$ goes
around the origin once, indicating the nontrivialness of the subchiral symmetry operator.

Now let us proceed to derive the low-energy Hamiltonians describing the gapless Dirac cones
on the $z$-normal surfaces. The methods are well developed~\cite{shen2013topological}. As usual, the first step is
to do a low-energy expansion of the bulk Hamiltonian around
the band-inversion momentum and decompose the Hamiltonian into two parts~\cite{Yan2018hosc},
i.e., $\mathcal{H}=\mathcal{H}_{0}+\mathcal{H}_{1}$,
with (see more details in the Supplemental Material~\cite{supplemental})
\begin{eqnarray}
\mathcal{H}_{0}(\bk)&=&[M(\bk_{s})+\frac{t_{z}}{2}k_{z}^{2}]\sigma_{z}s_{0}+\gamma(\bk_{s})k_{z}\sigma_{x}\tilde{s}_{x},\nonumber\\
\mathcal{H}_{1}(\bk)&=&\lambda(k_{x}\sigma_{x}\tilde{s}_{z}-k_{y}\sigma_{y}s_{0}),
\end{eqnarray}
where $M(\bk_{s})=m-2t-t_{z}+t(k_{x}^{2}+k_{y}^{2})/2$, and $\gamma(\bk_{s})=\frac{1}{2}\sqrt{\eta_{1}^{2}(k_{x}^{2}-k_{y}^{2})^{2}+4\eta_{2}^{2}k_{x}^{2}k_{y}^{2}}$.
Considering a half-infinity system occupying $z\geq0$ ($z\leq0$), replacing
$k_{z}\rightarrow -i\partial_{z}$, and solving the eigenvalue equation
$\mathcal{H}_{0}(\bk_{s},-i\partial_{z})\psi_{\alpha}(x,y,z)=0$ under the boundary conditions
$\psi_{\alpha}(z=0)=0$ and $\psi_{\alpha}(z\rightarrow\infty)=0$ ($\psi_{\alpha}(z\rightarrow-\infty)=0$), one will obtain
two solutions corresponding to the zero-energy boundary states at the bottom (top) surface. Their explicit forms
read~\cite{yan2016tunable}
\begin{eqnarray}
\psi_{\alpha}^{a}(x,y,z)=\mathcal{N}\sin(\kappa_{1}z)e^{-\kappa_{2}|z|}e^{i(k_{x}x+k_{y}y)}\chi_{\alpha}^{a},
\end{eqnarray}
where the superscript $a=\{t,b\}$ labels the top and bottom surfaces,
$\kappa_{1}=\sqrt{-2t_{z}M(\bk_{s})-\gamma^{2}(\bk_{s})}/t_{z}$, $\kappa_{2}=\gamma(\bk_{s})/t_{z}$,
$\mathcal{N}$ is a normalization constant, $\chi_{\alpha}^{t}$ satisfy
$\sigma_{y}\tilde{s}_{x}\chi_{\alpha}^{t}=\chi_{\alpha}^{t}$, and $\chi_{\alpha}^{b}$
satisfy $\sigma_{y}\tilde{s}_{x}\chi_{\alpha}^{b}=-\chi_{\alpha}^{b}$. The normalizability
of the wave functions determines the region hosting boundary states, which turns out
to be the region bound by the projection of the band-inversion surface, i.e., $M(\bk_{s})<0$.
Noteworthily, the point $\bk_{s}=\mathbf{0}$, however, needs to be excluded
since $\gamma(\bk_{s})$ vanishes at this point. This result is consistent with
the fact that the effective 1D Hamiltonian $\mathcal{H}(0,0,k_{z})$ is gapless, and
the projections of the two bulk Dirac nodes are exactly located at this surface time-reversal
invariant momentum~\cite{wang2013three}.

The low-energy Hamiltonians for the top and bottom surfaces are obtained by projecting $\mathcal{H}_{1}(\bk)$ onto the
Hilbert space spanned by the corresponding two zero-energy eigenstates. Since $[\sigma_{y}\tilde{s}_{x},\mathcal{C}]=0$,
we can choose $\chi_{\alpha}^{t/b}$ to be the eigenstates of the subchiral symmetry operator. Without
loss of generality, we choose $\chi_{\pm}^{t}=(|\sigma_{y}=1,\tilde{s}_{x}=1\rangle\pm |\sigma_{y}=-1,\tilde{s}_{x}=-1\rangle)/\sqrt{2}$,
and $\chi_{\pm}^{b}=(|\sigma_{y}=1,\tilde{s}_{x}=-1\rangle\mp |\sigma_{y}=-1,\tilde{s}_{x}=1\rangle)/\sqrt{2}$, so
that $\mathcal{C}\chi_{\pm}^{t/b}=\pm\chi_{\pm}^{t/b}$. Here
$|\sigma_{y}=\pm1,\tilde{s}_{x}=\pm1\rangle$ stands for $|\sigma_{y}=\pm1\rangle\otimes|\tilde{s}_{x}=\pm1\rangle$,
with $\sigma_{y}|\sigma_{y}=\pm1\rangle=\pm|\sigma_{y}=\pm1\rangle$ and $\tilde{s}_{x}|\tilde{s}_{x}=\pm1\rangle=\pm|\tilde{s}_{x}=\pm1\rangle$.
Accordingly, in the basis of $(\psi_{-}^{t},\psi_{+}^{t})^{T}$ or $(\psi_{+}^{b},\psi_{-}^{b})^{T}$,
the low-energy surface
Hamiltonians are found to take the {\zhongbo off-diagonal} form
\begin{eqnarray}
\mathcal{H}_{t/b}(\bk_{s})=\lambda(k_{x}\rho_{y}-k_{y}\rho_{x}),\label{surface}
\end{eqnarray}
where $\rho_{i}$ denote Pauli matrices  acting on the two eigenstates of the subchiral
symmetry operator. Apparently, the surface Hamiltonians take the exactly same form
as in TIs~\cite{Zhang2009bise}. However, here the linearly dispersive
SDCs have two fundamental differences. First, as discussed above,
surface states are absent at $\bk_{s}=\mathbf{0}$. This fact indicates the absence of Dirac node in
this class of SDCs. Second, here the basis functions are the eigenstates
of the subchiral symmetry operator, which themselves carry nontrivial topological
properties as the subchiral symmetry operator displays a nontrivial winding
with respect to the momentum. As will be shown below, this property has nontrivial
effects on the spin texture and Berry phase.

\begin{figure}[t]
\centering
\includegraphics[width=0.46\textwidth]{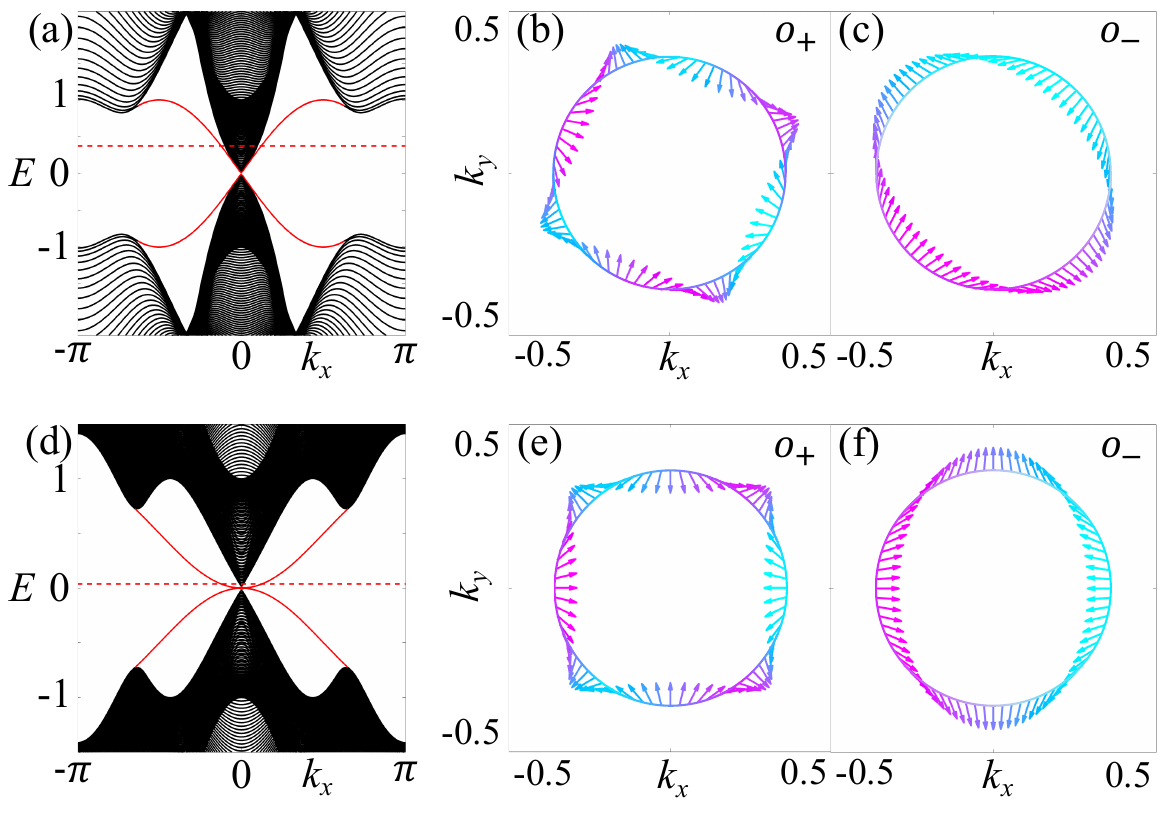}
\caption{(color online) Energy spectra at $k_{y}=0$ for a sample
with open (periodic) boundary conditions in the $z$ ($x$ and $y$) direction
and spin textures of the top-surface Dirac cones. The solid red lines in (a) and (d)
show the existence of linear and quadratic SDCs in the opposite-parity and same-parity
DSMs, respectively. The orbital-resolved spin textures in (b) and (c) [(e) and (f)]
are plotted on the iso-energy contour of the SDC illustrated by the red dashed line corresponding to
$E=0.375$ ($0.038$) in
(a) [(d)]. Common parameters are $m=3$, $t=t_{z}=2$, $\lambda=1$.  $\eta_{1}=\eta_{2}=5$ in (a)-(c),
and $\eta_{1}=\eta_{2}=0.5$ in (d)-(f).  }
\label{texture}
\end{figure}

To determine the spin texture and Berry phase, we need to first determine the
spinor part of the wave functions for the SDCs. To be specific, let us focus on the upper
band of the top-surface Dirac cone for a detailed discussion (the spin texture
for the bottom-surface Dirac cone is just opposite, and the Berry phase
is same). According to the form of
$\mathcal{H}_{t}$ in Eq.(\ref{surface}), it is easy to find that the eigenstate
for the upper band is $(1,ie^{i\phi_{\bk_{s}}})^{T}/\sqrt{2}$.
By further taking into account the nontrivial basis functions,
the corresponding spinor takes the  form
\begin{eqnarray}
|u(\bk_{s})\rangle=\frac{1}{\sqrt{2}}(\chi_{-}^{t}+ie^{i\phi_{\bk_{s}}}\chi_{+}^{t}).
\end{eqnarray}
Because the spin and orbital are entangled by spin-orbit coupling, we consider the
orbital-resolved spin texture\cite{Cao2013,Park2012texture,Zhang2013texture}, which are
given by
$\bar{s}_{i}^{(o_{\pm})}(\bk_{s})=\langle u(\bk_{s})|(\sigma_{0}\pm\sigma_{z})s_{i}|u(\bk_{s})\rangle/2$,
where the two superscripts $o_{+}$ and $o_{-}$ label the two orbitals (here
we ignore the constant factor $\hbar/2$ connecting the Pauli matrices
to the spin operators). A straightforward calculation obtains
\begin{eqnarray}
\bar{s}_{x}^{(o_{\pm})}(\bk_{s})&=&\pm[\sin(\theta_{\bk_{s}}\mp\phi_{\bk_{s}})]/2,\nonumber\\
\bar{s}_{y}^{(o_{\pm})}(\bk_{s})&=&\mp[\cos(\theta_{\bk_{s}}\mp\phi_{\bk_{s}})]/2,\nonumber\\
\bar{s}_{z}^{(o_{\pm})}(\bk_{s})&=&0.
\end{eqnarray}
The spin polarizations are aligned in the surface plane. This is similar to the spin textures
of the SDCs in TIs. However, here a striking difference
is that the spin textures depend on two phase angles rather than one as in TIs~\cite{Zhang2013texture}. Particularly,
the angle $\theta_{\bk_{s}}$ originates from the subchiral symmetry, and will
change $4\pi$ when the polar angle of the surface momentum changes $2\pi$.
Due to the unique two-phase-angle dependence, the two orbital-resolved spin textures display
a remarkable property, namely, their spin polarizations wind
one and three times respectively when $\bk_{s}$ winds the origin once, as shown in
Figs.\ref{texture}(a)-(c). This is rather different from the TI for which only one time
of winding will exhibit~\cite{hasan2010,qi2011}.

Also based on $|u(\bk_{s})\rangle$, the Berry connection is given by~\cite{Xiao2010review}
\begin{eqnarray}
A_{\alpha}(\bk_{s})=-i\langle u(\bk_{s})|\partial_{k_{\alpha}}u(\bk_{s})\rangle=\frac{1}{2}\partial_{k_{\alpha}}(\theta_{\bk_{s}}+\phi_{\bk_{s}}),
\end{eqnarray}
where $\alpha=\{x,y\}$. Since $\theta_{\bk_{s}}$ will wind $4\pi$
and $\phi_{\bk_{s}}$ will wind $2\pi$ when $\bk_{s}$ winds $2\pi$,
it indicates
that one particle will accumulate a $\pi$ (mod $2\pi$) Berry phase
when it goes around the surface Fermi loop once. This important result indicates that the quantized $\pi$ Berry
phase remains intact even though the singular Dirac node is absent in the SDCs.

{\it Quadratic nodeless SDCs in the same-parity  DSM.---}Let
us move our attention to the same-parity  DSM. Also focusing
on a cubic-lattice realization, the minimal model is given by~\cite{Yang2014}
\begin{eqnarray}
\mathcal{H}(\bk)&=&(m-t\cos k_{x}-t\cos k_{y}-t_{z}\cos k_{z})\sigma_{z}s_{0}\nonumber\\
&&+\lambda\sin k_{z}(\sin k_{x}\sigma_{x}s_{0}-\sin k_{y}\sigma_{y}s_{z})\nonumber\\
&&+\eta_{1}(\bk_{s})\sigma_{y}s_{x}+\eta_{2}(\bk_{s})\sigma_{y}s_{y}.\label{typeII}
\end{eqnarray}
Without loss of generality, below we again consider all parameters to be positive, and $|m-2t|<t_{z}<m$
so that the two bulk Dirac nodes are also located at $\bk_{D,\pm}$. Similar to the first model,
this model also supports interesting gapless topological states on the side surfaces and hinges~\cite{Szabo2020}. However, much less
is known about the top and bottom surfaces. Below we explore the surface states on these two surfaces.

The first thing to note is that the Hamiltonian (\ref{typeII}) also has a subchiral symmetry, with the symmetry operator given by
\begin{eqnarray}
\tilde{\mathcal{C}}=\sin\phi_{\bk_{s}}\sigma_{x}s_{0}+\cos\phi_{\bk_{s}}\sigma_{y}s_{z}.
\end{eqnarray}
 Also using the continuum-model approach,
we find that the wave functions of surface states on the top and bottom surfaces are given by
\begin{eqnarray}
\tilde{\psi}_{\alpha}^{a}(x,y,z)=\tilde{\mathcal{N}}\sin(\tilde{\kappa}_{1}z)
e^{-\tilde{\kappa}_{2}|z|}e^{i(k_{x}x+k_{y}y)}\tilde{\chi}_{\alpha}^{a},\label{wave2}
\end{eqnarray}
where $\tilde{\kappa}_{1}=\sqrt{-2t_{z}M(\bk_{s})-\lambda^{2}\bk_{s}^{2}}/t_{z}$,
$\tilde{\kappa}_{2}=\lambda|\bk_{s}|/t_{z}$,
and $\tilde{\chi}_{\alpha}^{a}$
satisfy $\tilde{\mathcal{C}}\tilde{\chi}_{\alpha}^{a}=\alpha\tilde{\chi}_{\alpha}^{a}$ with
$\alpha=\pm$.
The normalizability of the wave functions also suggests that the
region hosting surface states corresponds to $M(\bk_{s})<0$
but with the point $\bk_{s}=\mathbf{0}$
excluded. Without loss of generality, we choose
$\tilde{\chi}_{\pm}^{t}=|\sigma_{\pm}=\pm1,s_{z}=\pm1\rangle$,
and $\tilde{\chi}_{\mp}^{b}=|\sigma_{\pm}=\mp1,s_{z}=\pm 1\rangle$,
where $\sigma_{\pm}=\sin\phi_{\bk_{s}}\sigma_{x}\pm \cos\phi_{\bk_{s}}\sigma_{y}$.
In the basis of
$(\tilde{\psi}_{+}^{t},\tilde{\psi}_{-}^{t})^{T}$ or $(\tilde{\psi}_{-}^{b},\tilde{\psi}_{+}^{b})^{T}$,
the low-energy surface
Hamiltonians are found to take the {\zhongbo off-diagonal} form
\begin{eqnarray}
\mathcal{H}_{t/b}(\bk_{s})=\pm\left(
                     \begin{array}{cc}
                       0 & \eta_{-}(\bk_{s})e^{i\phi_{\bk_{s}}} \\
                       \eta_{+}(\bk_{s})e^{-i\phi_{\bk_{s}}} & 0 \\
                     \end{array}
                   \right),\label{surface2}
\end{eqnarray}
where $+$ ($-$) refers to the top (bottom) surface, and
$\eta_{\pm}(\bk_{s})=-\frac{\eta_{1}}{2}(k_{x}^{2}-k_{y}^{2})\pm i \eta_{2}k_{x}k_{y}$. It is easy
to see that the energy dispersions of the surface Hamiltonian are given by
$E_{\pm}(\bk_{s})=\pm\sqrt{\eta_{+}(\bk_{s})\eta_{-}(\bk_{s})}$, which are quadratic
rather than linear, as shown in Fig.\ref{texture}(d). It is worth emphasizing that
the Dirac node is also absent for this class of quadratic SDCs.

Again let us focus on the upper
band of the top-surface Dirac cone for a discussion of its spin texture and Berry phase.
The corresponding spinor part of the wave function is found to take  the form
\begin{eqnarray}
|\tilde{u}(\bk_{s})\rangle
=\frac{1}{\sqrt{2}}(\tilde{\chi}_{+}^{t}+e^{i(\theta_{\bk_{s}}-\phi_{\bk_{s}})}\tilde{\chi}_{-}^{t}).
\end{eqnarray}
Based on $|\tilde{u}(\bk_{s})\rangle$,  one finds
\begin{eqnarray}
\bar{s}_{x}^{(o_{\pm})}(\bk_{s})&=&[\cos(\theta_{\bk_{s}}\mp\phi_{\bk_{s}})]/2,\nonumber\\
\bar{s}_{y}^{(o_{\pm})}(\bk_{s})&=&[\sin(\theta_{\bk_{s}}\mp\phi_{\bk_{s}})]/2,\nonumber\\
\bar{s}_{z}^{(o_{\pm})}(\bk_{s})&=&0.
\end{eqnarray}
The two orbital-resolved spin textures also depend on two phase angles and display different windings,
as shown in Figs.\ref{texture}(e) and \ref{texture}(f). The Berry connection
is given by
\begin{eqnarray}
A_{\alpha}(\bk_{s})&=&-i\langle \tilde{u}(\bk_{s})|\partial_{k_{\alpha}}\tilde{u}(\bk_{s})\rangle=\frac{1}{2}\partial_{k_{\alpha}}(\theta_{\bk_{s}}-\phi_{\bk_{s}}).\,\label{berry2}
\end{eqnarray}
Similarly, this result indicates that the particle will accumulate a $\pi$ (mod $2\pi$) Berry phase
when it goes around the surface Fermi loop once. This is a remarkable result since usually
a quadratic cone is accompanied with a zero (mod $2\pi$) Berry phase~\cite{Novoselov2006}. From Eq.(\ref{berry2}),
it is apparent that the $\pi$ Berry phase is attributed to $\phi_{\bk_{s}}$, indicating
its origin from the subchiral symmetry rather than the quadratic band structure.

\begin{figure}[t]
\centering
\includegraphics[width=0.45\textwidth]{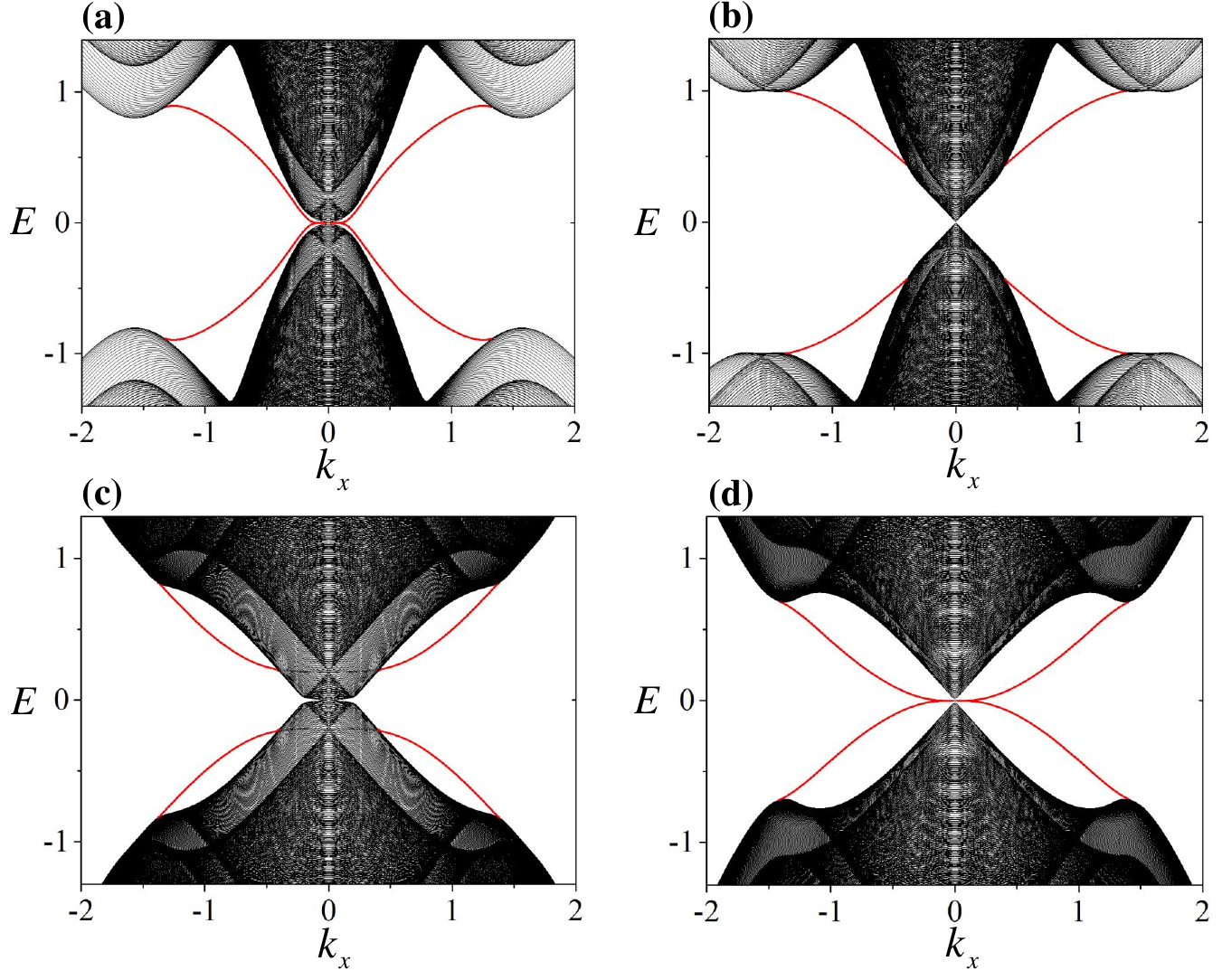}
\caption{(Color online) Energy spectra at $k_{y}=0$ for a sample with open boundary
conditions only in the $z$ direction (lattice sites $N_{z}=400$). Common parameters are $m=4$,
$t=t_{z}=2$,  and $\lambda=1$. The values of $(\eta_{1},\eta_{2},B_{1},B_{2})$ in (a), (b), (c)
and (d) are $(5,5,0.2,0)$,
$(5,5,0,0.2)$, $(1,1,0.2,0)$, and $(1,1,0,0.2)$, respectively. (a) and (b) correspond to the opposite-parity
DSM, and (c) and (d) refer to the same-parity DSM.
}\label{Zeeman}
\end{figure}

{\it Response to TRS-breaking fields.---}It is known that
the SDCs in TIs are protected by TRS, and the lift of
TRS can gap the SDCs~\cite{Qi2008TFT,Chen2010massive}. On the other hand, it is known that
TRS-breaking fields will split one bulk Dirac node into two Weyl nodes~\cite{Armitage2018}.
Here the SDCs are nodeless, and therefore are not protected by TRS.
To open a gap to the SDCs, mathematically a Dirac mass term of the form $m_{D}\rho_{z}$ is required
to enter into the surface Hamiltonian (\ref{surface}) or (\ref{surface2}). As
the basis functions for the surface Hamiltonians are eigenstates of the subchiral
symmetry operators, a necessary but not sufficient condition to generate the Dirac mass
term is that the TRS-breaking fields must commute with the subchiral symmetry operator.
To demonstrate the above arguments,
we consider two types of Zeeman splitting fields, i.e., $B_{1}\sigma_{0}s_{z}$ and $B_{2}\sigma_{z}s_{z}$.
For the opposite-parity DSM, $B_{1}\sigma_{0}s_{z}$ preserves the subchiral symmetry, while
$B_{2}\sigma_{z}s_{z}$ does not. The situation is just the opposite for the same-parity DSM.
As shown in Fig.\ref{Zeeman}, the results show that $B_{2}\sigma_{z}s_{z}$ gaps the SDCs in the opposite-parity DSM, while
$B_{1}\sigma_{0}s_{z}$ gaps the SDCs in the same-parity DSM, in consistent
with the above analysis. Interestingly, we note no matter whether the SDCs
are gapped or not, the surface states are always connected with the bulk nodes,
either at $E=0$ or $\pm B_{1}$ ($\pm B_{2}$), which can be viewed as a kind of
bulk-surface correspondence. Furthermore, for the gapless cases shown in Figs.\ref{Zeeman}(a) and \ref{Zeeman}(d),
we note that the Zeeman fields
flatten the SDCs, which may have nontrivial
interplay with interactions.

{\it Discussions and conclusions.---} We have unveiled the existence
of two types of nodeless SDCs with linear and quadratic dispersion,
quantized $\pi$ Berry phases and
unconventional spin textures, expanding our
understanding of the topological surface states and bulk-boundary correspondence
in DSMs. Our predictions
are of general relevance as our theory is based on two generic classes of
DSMs.  In experiments, the dispersion of the SDCs and the concomitant
unconventional spin textures can be directly detected  by using spin-resolved and
angle-resolved photoemission spectroscopy~\cite{zhang2019multiple,Souma2011texture,
Wang2011texture,Lv2015texture,Xu2016texture,Sakano2020texture,Dai2021texture}.
To conclude, {\zb our work exemplifies that the subchiral symmetry
can enrich the properties of the topological boundary states,}
and our findings diversify the types of SDCs with
fascinating properties,
opening directions for future studies of unconventional Dirac physics.

{\it Acknowledgements.---}We would like to thank Zhigang Wu for helpful
discussions and insightful comments. This work is supported by the
National Natural Science Foundation of China (Grant No.12174455),
the Natural Science Foundation of Guangdong Province
(Grant No. 2021B1515020026), {\zb and
the Guangdong Basic and Applied Basic Research Foundation
(Grant No. 2023B1515040023)}.

\bibliography{dirac}

\begin{widetext}
\clearpage
\begin{center}
\textbf{\large Supplemental Material for ``Anomalous Linear and Quadratic Nodeless Surface Dirac Cones \\in Three-Dimensional Dirac Semimetals''}\\
\vspace{4mm}
{Dongling Liu$^{1,*}$, Xiao-Jiao Wang$^{1,*}$, Yijie Mo$^{1}$, Zhongbo Yan$^{1,\dag}$}\\
\vspace{2mm}
{\em $^1$Guangdong Provincial Key Laboratory of Magnetoelectric Physics and Devices,\\
State Key Laboratory of Optoelectronic Materials and Technologies,\\
and School of Physics, Sun Yat-sen University, Guangzhou 510275, China}\\
\end{center}

\setcounter{equation}{0}
\setcounter{figure}{0}
\setcounter{table}{0}
\makeatletter
\renewcommand{\theequation}{S\arabic{equation}}
\renewcommand{\thefigure}{S\arabic{figure}}
\renewcommand{\bibnumfmt}[1]{[S#1]}

The supplemental material contains two sections. In the first section, we provide the derivation details of
the low-energy surface Hamiltonians, spin textures, and Berry phases for the opposite-parity Dirac semimetal.
In the second section, we derive the orbital-resolved spin textures associated with the surface Dirac cone in
a topological insulator to highlight the difference of the surface Dirac cones in Dirac semimetals and topological
insulators.

\section{I. The low-energy surface Hamiltonians, spin textures, and Berry phases for the  Dirac semimetals}

Because the derivation steps are rather similar for the two representative types of Dirac semimetals, here we restrict ourselves to
the opposite-parity Dirac semimetal for a detailed discussion.
We start from the tight-binding Hamiltonian describing the opposite-parity Dirac semimetal,
\begin{eqnarray}
\mathcal{H}(\bk)&=&(m-t\cos k_{x}-t\cos k_{y}-t_{z}\cos k_{z})\sigma_{z}s_{0}+\lambda (\sin k_{x}\sigma_{x}s_{z}-\sin k_{y}\sigma_{y}s_{0})\nonumber\\
&&+\eta_{1}(\cos k_{x}-\cos k_{y})\sin k_{z}\sigma_{x}s_{x}+\eta_{2}\sin k_{x}\sin k_{y}\sin k_{z}\sigma_{x}s_{y},\label{Hamiltonian1}
\end{eqnarray}
where $\sigma_{i}$ and $s_{i}$ are Pauli matrices respectively acting on orbital and spin degrees of freedom,
and $\sigma_{0}$ and $s_{0}$ are the corresponding $2\times2$ identity matrices. For notational simplicity, the lattice constants are set to unity.

Without loss of generality, we consider that the band inversion occurs at the center of the bulk Brillouin zone,
i.e., the time-reversal invariant momentum $\boldsymbol{\Gamma}=(0,0,0)$,
so that the bulk Dirac points are located on the $k_{z}$ axis going through $\boldsymbol{\Gamma}$. Because the surface states
originate from the band inversion, we expand the tight-binding Hamiltonian around the band inversion momentum
and obtain the continuum Hamiltonian to derive the low-energy Hamiltonians describing the surface states.
For each term in the Hamiltonian, we only keep the leading-order momentum terms to simplify
the analytical derivation. Accordingly, the continuum
Hamiltonian is given by
\begin{eqnarray}
\mathcal{H}(\bk)&=&[M(\bk_{s})+\frac{t_{z}}{2}k_{z}^{2}]\sigma_{z}s_{0}+\lambda(k_{x}\sigma_{x}s_{z}-k_{y}\sigma_{y}s_{0})\nonumber\\
&&-\frac{\eta_{1}}{2}(k_{x}^{2}-k_{y}^{2})k_{z}\sigma_{x}s_{x}+\eta_{2}k_{x}k_{y}k_{z}\sigma_{x}s_{y},
\end{eqnarray}
where $\bk_{s}=(k_{x},k_{y})$ denotes the momentum vector parallel to the $z$-normal surfaces,
and $M(\bk_{s})=m-2t-t_{z}+t(k_{x}^{2}+k_{y}^{2})/2$. Next, we define a new set of Pauli matrices
of the form
\begin{eqnarray}
&&\tilde{s}_{x}=\cos\theta_{\bk_{s}}s_{x}+\sin\theta_{\bk_{s}}s_{y},\nonumber\\
&&\tilde{s}_{y}=-\sin\theta_{\bk_{s}}s_{x}+\cos\theta_{\bk_{s}}s_{y},\nonumber\\
&&\tilde{s}_{z}=s_{z},
\end{eqnarray}
where $\theta_{\bk_{s}}=\arg[-\frac{\eta_{1}}{2}(k_{x}^{2}-k_{y}^{2})+i\eta_{2}k_{x}k_{y}]$, which is  well defined
except at $\bk_{s}=\mathbf{0}$. It is easy to check that the new set of Pauli matrices $\tilde{s}_{i}$ satisfies
the algebraic properties associated with the conventional Pauli matrices,
i.e., $[\tilde{s}_{i},\tilde{s}_{j}]=2i\epsilon_{ijk}\tilde{s}_{k}$ and $\{\tilde{s}_{i},\tilde{s}_{j}\}=2\delta_{ij}s_{0}$
for $i,j\in\{x,y,z\}$.
Using this new set of Pauli matrices, the continuum Hamiltonian can be rewritten as
\begin{eqnarray}
\mathcal{H}(\bk)=[M(\bk_{s})+\frac{t_{z}}{2}k_{z}^{2}]\sigma_{z}s_{0}+\lambda(k_{x}\sigma_{x}\tilde{s}_{z}-k_{y}\sigma_{y}s_{0})
+\gamma(\bk_{s})k_{z}\sigma_{x}\tilde{s}_{x},
\end{eqnarray}
where $\gamma(\bk_{s})=\frac{1}{2}\sqrt{\eta_{1}^{2}(k_{x}^{2}-k_{y}^{2})^{2}+4\eta_{2}^{2}k_{x}^{2}k_{y}^{2}}$. In this form,
it is easy to find that the Hamiltonian anticommutes with the operator $\mathcal{C}=\sigma_{x}\tilde{s}_{y}$. Conventionally,
when there exists a momentum-independent unitary operator anticommuting with the Hamiltonian, it implies that the Hamiltonian has
chiral symmetry. The momentum independence is because the chiral symmetry is a non-spatial symmetry. However, here the operator
$\mathcal{C}$ depends on momentum, thereby the anticommutation relation between $\mathcal{C}$ and $\mathcal{H}(\bk)$ does not suggest
the existence of the chiral symmetry. Nevertheless, here $\mathcal{C}$ only depends on $k_{x}$ and $k_{y}$. For situations
where $k_{x}$ and $k_{y}$ are good quantum numbers and can be viewed as a parameter, such as the surface states on the
$z$-normal surfaces considered below, this anticommutation relation has the conventional meaning of chiral symmetry.
In Ref.\cite{Mo2023}, the authors introduced the concept termed ``subchiral symmetry'' to
describe the existence of such an anticommutation relation, and showed that
the existence of subchiral symmetry can have nontrivial effects on the spin textures
and Berry phases of the boundary states.

Now let us derive the low-energy Hamiltonians describing the surface states on the $z$-normal surface states. To simplify
the derivation, we consider a half-infinity system to avoid the coupling between the surface states on the top and bottom
surfaces. First, we assume that the system occupies the region with coordinates satisfying $z\geq0$, accordingly the surface at $z=0$ corresponds to
the bottom surface. Second, we decompose the continuum Hamiltonian into two parts~\cite{Yan2018hosc},
i.e., $\mathcal{H}=\mathcal{H}_{0}+\mathcal{H}_{1}$, where
\begin{eqnarray}
\mathcal{H}_{0}(\bk)&=&[M(\bk_{s})+\frac{t_{z}}{2}k_{z}^{2}]\sigma_{z}s_{0}+\gamma(\bk_{s})k_{z}\sigma_{x}\tilde{s}_{x},\nonumber\\
\mathcal{H}_{1}(\bk)&=&\lambda(k_{x}\sigma_{x}\tilde{s}_{z}-k_{y}\sigma_{y}s_{0}).
\end{eqnarray}
In the presence of a boundary in the $z$ direction, the translation symmetry is broken in the $z$ direction, and
$k_{z}$ needs to be replaced by $-i\partial_{z}$. Accordingly,
\begin{eqnarray}
\mathcal{H}_{0}(\bk_{s},-i\partial_{z})=[M(\bk_{s})-\frac{t_{z}}{2}\partial_{z}^{2}]\sigma_{z}s_{0}
-i\gamma(\bk_{s})\partial_{z}\sigma_{x}\tilde{s}_{x},
\end{eqnarray}
and $\mathcal{H}_{1}$ retains its form as the translation symmetry converses in the $xy$ plane. The next
step is to solve the eigenvalue equation
\begin{eqnarray}
\mathcal{H}_{0}(\bk_{s},-i\partial_{z})\psi(x,y,z)=E(\bk_{s})\psi(x,y,z)
\end{eqnarray}
under the boundary conditions $\psi(x,y,z=0)=\psi(x,y,z\rightarrow+\infty)=0$. We find that there
are two zero-energy solutions satisfying the boundary conditions, and the solutions have the form
\begin{eqnarray}
\psi_{\alpha}^{b}(x,y,z)=\mathcal{N}\sin(\kappa_{1}z)e^{-\kappa_{2}z}e^{i(k_{x}x+k_{y}y)}\chi_{\alpha}^{b},\label{wf}
\end{eqnarray}
where $\alpha=\pm$, $\kappa_{1}=\sqrt{-2t_{z}M(\bk_{s})-\gamma^{2}(\bk_{s})}/t_{z}$, $\kappa_{2}=\gamma(\bk_{s})/t_{z}$,
$\mathcal{N}=2\sqrt{|\kappa_{2}(\kappa_{1}^{2}+\kappa_{2}^{2})/\kappa_{1}^{2}|}$ is the normalization constant, and $\chi_{\alpha}^{b}$ satisfy
$\sigma_{y}\tilde{s}_{x}\chi_{\alpha}^{b}=-\chi_{\alpha}^{b}$. To be normalizable,
the two parameters $\kappa_{1}$ and $\kappa_{2}$ in the wave functions need to satisfy the constraint
that $\kappa_{1}^{2}+\kappa_{2}^{2}>0$ and $\kappa_{2}\neq0$.  Accordingly,
one finds that the zero-energy surface states exist within the region
$M(\bk_{s})<0$ but with the special point $\bk_{s}=\mathbf{0}$ excluded.

From the equation $\sigma_{y}\tilde{s}_{x}\chi_{\alpha}^{b}=-\chi_{\alpha}^{b}$, it is easy to see
that the spinors $\chi_{\alpha}^{b}$ can be chosen as $|\sigma_{y}=1,\tilde{s}_{x}=-1\rangle$ and
$|\sigma_{y}=-1,\tilde{s}_{x}=1\rangle$. However, as $\sigma_{y}\tilde{s}_{x}$ commutes with the subchiral
symmetry operator $\mathcal{C}=\sigma_{x}\tilde{s}_{y}$, we choose $\chi_{\alpha}^{b}$ to
be simultaneously the eigenstates of $\sigma_{y}\tilde{s}_{x}$ and $\sigma_{x}\tilde{s}_{y}$. Without loss of generality,
we choose
\begin{eqnarray}
\chi_{+}^{b}&=&\frac{1}{\sqrt{2}}(|\sigma_{y}=1,\tilde{s}_{x}=-1\rangle-|\sigma_{y}=-1,\tilde{s}_{x}=1\rangle),\nonumber\\
\chi_{-}^{b}&=&\frac{1}{\sqrt{2}}(|\sigma_{y}=1,\tilde{s}_{x}=-1\rangle+|\sigma_{y}=-1,\tilde{s}_{x}=1\rangle).
\end{eqnarray}
One can check $\mathcal{C}\chi_{\pm}^{b}=\pm\chi_{\pm}^{b}$. The low-energy Hamiltonian for the surface states
is then obtained by projecting $\mathcal{H}_{1}$ onto the two-dimensional Hilbert space spanned by $\psi_{+}^{b}$ and $\psi_{-}^{b}$.
To be specific, we choose the basis to be $(\psi_{+}^{b},\psi_{-}^{b})^{T}$, then the matrix elements of the low-energy surface
Hamiltonian are given by
\begin{eqnarray}
&&[\mathcal{H}_{b}(\bk_{s})]_{11}=\int_{0}^{+\infty}
[\psi_{+}^{b}(x,y,z)]^{\dag}\mathcal{H}_{1}(\bk_{s},-i\partial_{z})\psi_{+}^{b}(x,y,z)dz,\nonumber\\
&&[\mathcal{H}_{b}(\bk_{s})]_{12}=\int_{0}^{+\infty}
[\psi_{+}^{b}(x,y,z)]^{\dag}\mathcal{H}_{1}(\bk_{s},-i\partial_{z})\psi_{-}^{b}(x,y,z)dz,\nonumber\\
&&[\mathcal{H}_{b}(\bk_{s})]_{21}=\int_{0}^{+\infty}
[\psi_{-}^{b}(x,y,z)]^{\dag}\mathcal{H}_{1}(\bk_{s},-i\partial_{z})\psi_{+}^{b}(x,y,z)dz,\nonumber\\
&&[\mathcal{H}_{b}(\bk_{s})]_{22}=\int_{0}^{+\infty}
[\psi_{-}^{b}(x,y,z)]^{\dag}\mathcal{H}_{1}(\bk_{s},-i\partial_{z})\psi_{-}^{b}(x,y,z)dz.
\end{eqnarray}
Now $\bk_{s}=(k_{x},k_{y})$ has the meaning of the momentum vector in the surface Brillouin zone. In the matrix form, one has
\begin{eqnarray}
\mathcal{H}_{b}(\bk_{s})=\lambda\left(
                               \begin{array}{cc}
                                 0 & -k_{y}-ik_{x} \\
                                 -k_{y}+ik_{x} & 0 \\
                               \end{array}
                             \right).
\end{eqnarray}
In terms of Pauli matrices, the low-energy surface Hamiltonian can be rewritten in the form
\begin{eqnarray}
\mathcal{H}_{b}(\bk_{s})=\lambda(k_{x}\rho_{y}-k_{y}\rho_{x}),\label{bottom}
\end{eqnarray}
where $\rho_{i}$ denote Pauli matrices. Because the underlying basis functions are
the eigenstates of the subchiral symmetry operator, one can see that the surface Hamiltonian takes an
off-diagonal form. From this surface Hamiltonian, one obtains the energy spectra of
the surface states, which  read
\begin{eqnarray}
E_{\pm}(\bk_{s})=\pm\lambda\sqrt{k_{x}^{2}+k_{y}^{2}}=\pm\lambda|\bk_{s}|.
\end{eqnarray}
The energy spectra form a linear Dirac cone. However, since the surface
states do not exist at $\bk_{s}=\mathbf{0}$, this linear surface Dirac cone does not
host the Dirac node occurring at this surface time-reversal invariant momentum,
which is fundamentally different from the linear surface Dirac cone
in topological insulators. The latter is known to host a time-reversal-symmetry-enforced
Dirac node in its structure.

To obtain the low-energy Hamiltonian describing the surface states on
the top surface, it is convenient to assume a half-infinity geometry with the region
$z\leq0$ occupied. Then the surface at $z=0$ corresponds to the top surface. Similarly,
the next step is to solve the eigenvalue equation
\begin{eqnarray}
\mathcal{H}_{0}(\bk_{s},-i\partial_{z})\psi(x,y,z)=E(\bk_{s})\psi(x,y,z),
\end{eqnarray}
but with the boundary conditions modified as $\psi(x,y,z=0)=\psi(x,y,z\rightarrow-\infty)=0$. It is
straightforward to find that there are also two zero-energy solutions satisfying
the boundary conditions, and the solutions have a form similar to that in Eq.(\ref{wf}).
Namely, we have
\begin{eqnarray}
\psi_{\alpha}^{t}(x,y,z)=\mathcal{N}\sin(\kappa_{1}z)e^{\kappa_{2}z}e^{i(k_{x}x+k_{y}y)}\chi_{\alpha}^{t}.\label{wf2}
\end{eqnarray}
Here the parameters $\kappa_{1}$, $\kappa_{2}$ and $\mathcal{N}$ are the same as before. The only difference
is that now the spinors $\chi_{\alpha}^{t}$ satisfy the eigenvalue equation,
$\sigma_{y}\tilde{s}_{x}\chi_{\alpha}^{t}=\chi_{\alpha}^{t}$. Two natural solutions
are $|\sigma_{y}=1,\tilde{s}_{x}=1\rangle$ and $|\sigma_{y}=-1,\tilde{s}_{x}=-1\rangle$.
However, since the subchiral symmetry operator commutes with $\sigma_{y}\tilde{s}_{x}$, again
we choose the eigenstates to be simultaneously the eigenstates of $\sigma_{y}\tilde{s}_{x}$
and the subchiral symmetry operator. Concretely, we choose
\begin{eqnarray}
\chi_{+}^{t}&=&\frac{1}{\sqrt{2}}(|\sigma_{y}=1,\tilde{s}_{x}=1\rangle+|\sigma_{y}=-1,\tilde{s}_{x}=-1\rangle),\nonumber\\
\chi_{-}^{t}&=&\frac{1}{\sqrt{2}}(|\sigma_{y}=1,\tilde{s}_{x}=1\rangle-|\sigma_{y}=-1,\tilde{s}_{x}=-1\rangle).
\end{eqnarray}
It is easy to check that $\mathcal{C}\chi_{\pm}^{t}=\pm\chi_{\pm}^{t}$. Choose the basis to be
$(\psi_{-}^{t},\psi_{+}^{t})^{T}$ and project $\mathcal{H}_{1}$ onto it, we obtain the low-energy
surface Hamiltonian for the top surface, which takes the exactly same form as Eq.(\ref{bottom}),
i.e.,
\begin{eqnarray}
\mathcal{H}_{t}(\bk_{s})=\lambda(k_{x}\rho_{y}-k_{y}\rho_{x}).\label{top}
\end{eqnarray}
Despite the same form, one needs to note that here the basis is different.
Eqs.(\ref{bottom}) and (\ref{top}) give the Eq.(6) in the main article.

In the following, let us derive the spin textures associated with the surface states.
Without loss of generality, we focus on the upper-half cone. We first consider the bottom surface.
According to the low-energy surface Hamiltonian in Eq.(\ref{bottom}), one knows that the
eigenstate for the upper band is of the simple form $(1,ie^{i\phi_{\bk_{s}}})^{T}/\sqrt{2}$, where
$\phi_{\bk_{s}}=\text{arg}(k_{x}+ik_{y})$. However,
here the basis is not the usual spin-up and spin-down basis, thereby one has to take into account the basis
functions to determine the spin textures correctly. In addition, since both the spin texture and the Berry phase are determined by the
spinor part of the wave function, we can ignore the space dependence and just focus on the spinor part describing
the internal degrees of freedom.
By taking into account the basis functions, the spinor for the upper band of the surface states on the bottom
surface is given by
\begin{eqnarray}
|u^{b}(\bk_{s})\rangle=\frac{1}{\sqrt{2}}(\chi_{+}^{b}+ie^{i\phi_{\bk_{s}}}\chi_{-}^{b}).
\end{eqnarray}
Because the spin and orbital are strongly coupled by spin-orbit coupling, we should consider the orbital-resolved
spin textures, which are given by
\begin{eqnarray}
\bar{s}_{i}^{b,o_{\pm}}(\bk_{s})=\langle u^{b}(\bk_{s})|\frac{(\sigma_{0}\pm\sigma_{z})}{2}s_{i}|u^{b}(\bk_{s})\rangle.
\end{eqnarray}
where the two superscripts $o_{+}$ and $o_{-}$ label the two orbitals (here
we ignore the constant factor $\hbar/2$ connecting the Pauli matrices
to the spin operators).

Before proceeding, it is worth noting that since $\tilde{s}_{x}=\cos\theta_{\bk_{s}}s_{x}+\sin\theta_{\bk_{s}}s_{y}$, the explicit expressions of the two eigenstates
of $\tilde{s}_{x}$ are
\begin{eqnarray}
|\tilde{s}_{x}=1\rangle=\frac{1}{\sqrt{2}}\left(
                                            \begin{array}{c}
                                              1 \\
                                              e^{i\theta_{\bk_{s}}} \\
                                            \end{array}
                                          \right),\quad |\tilde{s}_{x}=-1\rangle=\frac{1}{\sqrt{2}}\left(
                                            \begin{array}{c}
                                              1 \\
                                              -e^{i\theta_{\bk_{s}}} \\
                                            \end{array}
                                          \right).
\end{eqnarray}
It is easy to check that the above choice maintains the usual property of Pauli matrices, namely,
\begin{eqnarray}
&&\tilde{s}_{y}|\tilde{s}_{x}=1\rangle=-i|\tilde{s}_{x}=-1\rangle,\quad \tilde{s}_{y}|\tilde{s}_{x}=-1\rangle=i|\tilde{s}_{x}=1\rangle,\nonumber\\
&&\tilde{s}_{z}|\tilde{s}_{x}=1\rangle=|\tilde{s}_{x}=-1\rangle,\quad \tilde{s}_{z}|\tilde{s}_{x}=-1\rangle=|\tilde{s}_{x}=1\rangle.
\end{eqnarray}

Using the following algebraic results,
\begin{eqnarray}
\langle \tilde{s}_{x}=\pm1|s_{x}|\tilde{s}_{x}=\pm1\rangle&=&\pm\frac{1}{2}(e^{i\theta_{\bk_{s}}}+e^{-i\theta_{\bk_{s}}})=\pm\cos\theta_{\bk_{s}},\nonumber\\
\langle \tilde{s}_{x}=\pm1|s_{x}|\tilde{s}_{x}=\mp1\rangle&=&\mp\frac{1}{2}(e^{i\theta_{\bk_{s}}}-e^{-i\theta_{\bk_{s}}})=\mp i\sin\theta_{\bk_{s}}
\nonumber\\
\langle \tilde{s}_{x}=\pm1|s_{y}|\tilde{s}_{x}=\pm1\rangle&=&\mp\frac{i}{2}(e^{i\theta_{\bk_{s}}}-e^{-i\theta_{\bk_{s}}})=\pm\sin\theta_{\bk_{s}},\nonumber\\
\langle \tilde{s}_{x}=\pm1|s_{y}|\tilde{s}_{x}=\mp1\rangle&=&\pm\frac{i}{2}(e^{i\theta_{\bk_{s}}}+e^{-i\theta_{\bk_{s}}})=\pm i\cos\theta_{\bk_{s}}\nonumber\\
\langle \tilde{s}_{x}=\pm1|s_{z}|\tilde{s}_{x}=\pm1\rangle&=&0,\nonumber
\end{eqnarray}
it is straightforward to find
\begin{eqnarray}
\bar{s}_{x}^{b,0}(\bk_{s})&=&\langle u^{b}(\bk_{s})|\sigma_{0}s_{x}|u^{b}(\bk_{s})\rangle\nonumber\\
&=&\frac{1}{2}[(\chi_{+}^{b})^{\dag}\sigma_{0}s_{x}\chi_{+}^{b}+(\chi_{-}^{b})^{\dag}\sigma_{0}s_{x}\chi_{-}^{b}]
+\frac{1}{2}[-ie^{-i\phi_{\bk_{s}}}(\chi_{-}^{b})^{\dag}\sigma_{0}s_{x}\chi_{+}^{b}+ie^{i\phi_{\bk_{s}}}
(\chi_{+}^{b})^{\dag}\sigma_{0}s_{x}\chi_{-}^{b}]\nonumber\\
&=&\frac{1}{2}\sum_{\alpha=\pm1}\langle\tilde{s}_{x}=\alpha|s_{x}|\tilde{s}_{x}=\alpha\rangle
-\frac{i}{4}e^{-i\phi_{\bk_{s}}}(\langle \tilde{s}_{x}=-1|s_{x}|\tilde{s}_{x}=-1\rangle
-\langle \tilde{s}_{x}=1|s_{x}|\tilde{s}_{x}=1\rangle)\nonumber\\
&&+\frac{i}{4}e^{i\phi_{\bk_{s}}}(\langle \tilde{s}_{x}=-1|s_{x}|\tilde{s}_{x}=-1\rangle
-\langle \tilde{s}_{x}=1|s_{x}|\tilde{s}_{x}=1\rangle)\nonumber\\
&=&\sin\phi_{\bk_{s}}\cos\theta_{\bk_{s}},
\end{eqnarray}
and
\begin{eqnarray}
\bar{s}_{x}^{b,z}(\bk_{s})&=&\langle u^{b}(\bk_{s})|\sigma_{z}s_{x}|u^{b}(\bk_{s})\rangle\nonumber\\
&=&\frac{1}{2}[(\chi_{+}^{b})^{\dag}\sigma_{z}s_{x}\chi_{+}^{b}+(\chi_{-}^{b})^{\dag}\sigma_{z}s_{x}\chi_{-}^{b}]
+\frac{1}{2}[-ie^{-i\phi_{\bk_{s}}}(\chi_{-}^{b})^{\dag}\sigma_{z}s_{x}\chi_{+}^{b}+ie^{i\phi_{\bk_{s}}}
(\chi_{+}^{b})^{\dag}\sigma_{z}s_{x}\chi_{-}^{b}]\nonumber\\
&=&-\frac{i}{4}e^{-i\phi_{\bk_{s}}}(\langle \tilde{s}_{x}=1|s_{x}|\tilde{s}_{x}=-1\rangle
-\langle \tilde{s}_{x}=-1|s_{x}|\tilde{s}_{x}=1\rangle)\nonumber\\
&&+\frac{i}{4}e^{i\phi_{\bk_{s}}}(\langle \tilde{s}_{x}=-1|s_{x}|\tilde{s}_{x}=1\rangle
-\langle \tilde{s}_{x}=1|s_{x}|\tilde{s}_{x}=-1\rangle)\nonumber\\
&=&-\cos\phi_{\bk_{s}}\sin\theta_{\bk_{s}}.
\end{eqnarray}
Similar calculations show that
\begin{eqnarray}
\bar{s}_{y}^{b,0}(\bk_{s})&=&\frac{1}{2}\sum_{\alpha=\pm1}\langle\tilde{s}_{x}=\alpha|s_{y}|\tilde{s}_{x}=\alpha\rangle
-\frac{i}{4}e^{-i\phi_{\bk_{s}}}(\langle \tilde{s}_{x}=-1|s_{y}|\tilde{s}_{x}=-1\rangle
-\langle \tilde{s}_{x}=1|s_{y}|\tilde{s}_{x}=1\rangle)\nonumber\\
&&+\frac{i}{4}e^{i\phi_{\bk_{s}}}(\langle \tilde{s}_{x}=-1|s_{y}|\tilde{s}_{x}=-1\rangle
-\langle \tilde{s}_{x}=1|s_{y}|\tilde{s}_{x}=1\rangle)\nonumber\\
&=&\sin\phi_{\bk_{s}}\sin\theta_{\bk_{s}},\nonumber\\
\bar{s}_{y}^{b,z}(\bk_{s})&=&
-\frac{i}{4}e^{-i\phi_{\bk_{s}}}(\langle \tilde{s}_{x}=1|s_{y}|\tilde{s}_{x}=-1\rangle
-\langle \tilde{s}_{x}=-1|s_{y}|\tilde{s}_{x}=1\rangle)\nonumber\\
&&+\frac{i}{4}e^{i\phi_{\bk_{s}}}(\langle \tilde{s}_{x}=-1|s_{y}|\tilde{s}_{x}=1\rangle
-\langle \tilde{s}_{x}=1|s_{y}|\tilde{s}_{x}=-1\rangle)\nonumber\\
&=&\cos\phi_{\bk_{s}}\cos\theta_{\bk_{s}},\nonumber\\
\bar{s}_{z}^{b,0}(\bk_{s})&=&\bar{s}_{z}^{b,z}(\bk_{s})=0.
\end{eqnarray}
Accordingly, we have
\begin{eqnarray}
\bar{s}_{x}^{b,o_{+}}(\bk_{s})&=&\frac{1}{2}(\bar{s}_{x}^{b,0}(\bk_{s})+\bar{s}_{x}^{b,z}(\bk_{s}))
=-\frac{1}{2}\sin(\theta_{\bk_{s}}-\phi_{\bk_{s}}),\nonumber\\
\bar{s}_{y}^{b,o_{+}}(\bk_{s})&=&\frac{1}{2}(\bar{s}_{y}^{b,0}(\bk_{s})+\bar{s}_{y}^{b,z}(\bk_{s}))
=\frac{1}{2}\cos(\theta_{\bk_{s}}-\phi_{\bk_{s}}),\nonumber\\
\bar{s}_{z}^{b,o_{+}}(\bk_{s})&=&0,\nonumber\\
\end{eqnarray}
and
\begin{eqnarray}
\bar{s}_{x}^{b,o_{-}}(\bk_{s})&=&\frac{1}{2}(\bar{s}_{x}^{b,0}(\bk_{s})-\bar{s}_{x}^{b,z}(\bk_{s}))
=\frac{1}{2}\sin(\theta_{\bk_{s}}+\phi_{\bk_{s}}),\nonumber\\
\bar{s}_{y}^{b,o_{-}}(\bk_{s})&=&\frac{1}{2}(\bar{s}_{y}^{b,0}(\bk_{s})-\bar{s}_{y}^{b,z}(\bk_{s}))
=-\frac{1}{2}\cos(\theta_{\bk_{s}}+\phi_{\bk_{s}}),\nonumber\\
\bar{s}_{z}^{b,o_{-}}(\bk_{s})&=&0.
\end{eqnarray}

For the top surface, similar calculations show that the spin texture for the upper-half cone is just the opposite, i.e.,
\begin{eqnarray}
&&\bar{s}_{x}^{t,0}(\bk_{s})=-\sin\phi_{\bk_{s}}\cos\theta_{\bk_{s}}, \quad
\bar{s}_{y}^{t,0}(\bk_{s})=-\sin\phi_{\bk_{s}}\sin\theta_{\bk_{s}},\quad
\bar{s}_{z}^{t,0}(\bk_{s})=0;\nonumber\\
&&\bar{s}_{x}^{t,z}(\bk_{s})=\cos\phi_{\bk_{s}}\sin\theta_{\bk_{s}}, \quad
\bar{s}_{y}^{t,z}(\bk_{s})=-\cos\phi_{\bk_{s}}\cos\theta_{\bk_{s}},\quad
\bar{s}_{z}^{t,z}(\bk_{s})=0.
\end{eqnarray}
Accordingly, we have
\begin{eqnarray}
&&\bar{s}_{x}^{t,o_{+}}(\bk_{s})=\frac{1}{2}\sin(\theta_{\bk_{s}}-\phi_{\bk_{s}}), \quad
\bar{s}_{y}^{t,o_{+}}(\bk_{s})=-\frac{1}{2}\cos(\theta_{\bk_{s}}-\phi_{\bk_{s}}),\quad
\bar{s}_{z}^{t,o_{+}}(\bk_{s})=0;\nonumber\\
&&\bar{s}_{x}^{t,o_{-}}(\bk_{s})=-\frac{1}{2}\sin(\theta_{\bk_{s}}+\phi_{\bk_{s}}), \quad
\bar{s}_{y}^{t,o_{-}}(\bk_{s})=\frac{1}{2}\cos(\theta_{\bk_{s}}+\phi_{\bk_{s}}),\quad
\bar{s}_{z}^{t,o_{-}}(\bk_{s})=0.
\end{eqnarray}
For each orbital, the spin textures for the top and bottom surfaces are opposite, which can simply be inferred from the fact that
the top and bottom surfaces are related by a mirror reflection ($\mathcal{M}_{z}$) that reverses the in-plane
spin polarizations.

Now we proceed to determine the Berry connection and Berry phase associated with the surface states.
The Berry connection for the surface states on the bottom surface is given by
\begin{eqnarray}
A_{\alpha}^{b}(\bk_{s})=-i\langle u^{b}(\bk_{s})|\partial_{k_{\alpha}}u^{b}(\bk_{s})\rangle,
\end{eqnarray}
where $\alpha=\{x,y\}$. By using the following results,
\begin{eqnarray}
&&-i(\chi_{+}^{b})^{\dag}\partial_{k_{\alpha}}\chi_{+}^{b}=-\frac{i}{2}\sum_{\alpha=\pm1}\langle \tilde{s}_{x}=\alpha|\partial_{k_{\alpha}}|\tilde{s}_{x}=\alpha\rangle=\frac{1}{2}\partial_{k_{\alpha}}\theta_{\bk_{s}},\nonumber\\
&&-i(ie^{i\phi_{\bk_{s}}}\chi_{-}^{b})^{\dag}\partial_{k_{\alpha}}(ie^{i\phi_{\bk_{s}}}\chi_{-}^{b})=-\frac{i}{2}\sum_{\alpha=\pm1}\langle \tilde{s}_{x}=\alpha|\partial_{k_{\alpha}}|\tilde{s}_{x}=\alpha\rangle+\partial_{k_{\alpha}}\phi_{\bk_{s}}
=\frac{1}{2}\partial_{k_{\alpha}}\theta_{\bk_{s}}+\partial_{k_{\alpha}}\phi_{\bk_{s}},\nonumber\\
&&-i(\chi_{+}^{b})^{\dag}\partial_{k_{\alpha}}(ie^{i\phi_{\bk_{s}}}\chi_{-}^{b})=\frac{1}{2}e^{i\phi_{\bk_{s}}}\sum_{\alpha=\pm1}
\alpha\langle \tilde{s}_{x}=\alpha|\partial_{k_{\alpha}}|\tilde{s}_{x}=\alpha\rangle=0,\nonumber\\
&&-i(ie^{i\phi_{\bk_{s}}}\chi_{-}^{b})^{\dag}\partial_{k_{\alpha}}(\chi_{+}^{b})=-\frac{1}{2}e^{-i\phi_{\bk_{s}}}\sum_{\alpha=\pm1}
\alpha\langle \tilde{s}_{x}=\alpha|\partial_{k_{\alpha}}|\tilde{s}_{x}=\alpha\rangle=0,
\end{eqnarray}
one obtains
\begin{eqnarray}
A_{\alpha}^{b}(\bk_{s})=\frac{1}{2}(\partial_{k_{\alpha}}\theta_{\bk_{s}}+\partial_{k_{\alpha}}\phi_{\bk_{s}}).
\end{eqnarray}
The Berry phase along a closed loop encircling the center of the surface Brillouin zone (the
surface time-reversal invariant momentum at $\bk_{s}=\mathbf{0}$) is given by
\begin{eqnarray}
\phi_{B}=\oint_{l}\mathbf{A}_{\alpha}^{b}(\bk_{s})\cdot d\bk_{l}.
\end{eqnarray}
As $\phi_{\bk_{s}}$ and $\theta_{\bk_{s}}$ change $2\pi$ and $4\pi$ respectively when the polar angle
of the momentum vector $\bk_{s}$ changes $2\pi$, the above integral gives $\pi$ (mod $2\pi$). It is worth mentioning
that the Berry phase
is only gauge invariant in the sense of modulo $2\pi$.
This result suggests that the Berry phase on a surface Fermi loop takes a quantized value of
$\pi$ even though the Dirac node is absent in the surface Dirac cone.

For the upper-half cone of the top surface, the corresponding spinor is of the form
\begin{eqnarray}
|u^{t}(\bk_{s})\rangle=\frac{1}{\sqrt{2}}(\chi_{-}^{t}+ie^{i\phi_{\bk_{s}}}\chi_{+}^{t}).
\end{eqnarray}
Similar calculations give
\begin{eqnarray}
A_{\alpha}^{t}(\bk_{s})=-i\langle u^{t}(\bk_{s})|\partial_{k_{\alpha}}u^{t}(\bk_{s})\rangle=\frac{1}{2}(\partial_{k_{\alpha}}\theta_{\bk_{s}}+\partial_{k_{\alpha}}\phi_{\bk_{s}}).
\end{eqnarray}
The same form of the Berry connection suggests that the iso-energy contour of the top-surface Dirac cone is also
characterized by a quantized $\pi$ Berry phase.

For the same-parity Dirac semimetal, similar calculations give
\begin{eqnarray}
&&\bar{s}_{x}^{t,0}(\bk_{s})=\cos\phi_{\bk_{s}}\cos\theta_{\bk_{s}}, \quad
\bar{s}_{y}^{t,0}(\bk_{s})=\cos\phi_{\bk_{s}}\sin\theta_{\bk_{s}},\quad
\bar{s}_{z}^{t,0}(\bk_{s})=0;\nonumber\\
&&\bar{s}_{x}^{t,z}(\bk_{s})=\sin\phi_{\bk_{s}}\sin\theta_{\bk_{s}}, \quad
\bar{s}_{y}^{t,z}(\bk_{s})=-\sin\phi_{\bk_{s}}\cos\theta_{\bk_{s}},\quad
\bar{s}_{z}^{t,z}(\bk_{s})=0.
\end{eqnarray}
Therefore, for the two orbitals, we have
\begin{eqnarray}
\bar{s}_{x}^{t,o_{+}}(\bk_{s})&=&\frac{1}{2}(\bar{s}_{x}^{t,0}(\bk_{s})+\bar{s}_{x}^{t,z}(\bk_{s}))=\frac{1}{2}\cos(\theta_{\bk_{s}}-\phi_{\bk_{s}}),\nonumber\\
\bar{s}_{y}^{t,o_{+}}(\bk_{s})&=&\frac{1}{2}(\bar{s}_{y}^{t,0}(\bk_{s})+\bar{s}_{y}^{t,z}(\bk_{s}))=\frac{1}{2}\sin(\theta_{\bk_{s}}-\phi_{\bk_{s}}),\nonumber\\
\bar{s}_{z}^{t,o_{+}}(\bk_{s})&=&\frac{1}{2}(\bar{s}_{z}^{t,0}(\bk_{s})+\bar{s}_{z}^{t,z}(\bk_{s}))=0,\nonumber\\
\bar{s}_{x}^{t,o_{-}}(\bk_{s})&=&\frac{1}{2}(\bar{s}_{x}^{t,0}(\bk_{s})-\bar{s}_{x}^{t,z}(\bk_{s}))=\frac{1}{2}\cos(\theta_{\bk_{s}}+\phi_{\bk_{s}}),\nonumber\\
\bar{s}_{y}^{t,o_{-}}(\bk_{s})&=&\frac{1}{2}(\bar{s}_{y}^{t,0}(\bk_{s})-\bar{s}_{y}^{t,z}(\bk_{s}))=\frac{1}{2}\sin(\theta_{\bk_{s}}+\phi_{\bk_{s}}),\nonumber\\
\bar{s}_{z}^{t,o_{-}}(\bk_{s})&=&\frac{1}{2}(\bar{s}_{z}^{t,0}(\bk_{s})-\bar{s}_{z}^{t,z}(\bk_{s}))=0,
\end{eqnarray}
Similarly, the orbital-resolved spin textures for the bottom surface are just opposite.

\section{II. Orbital-resolved spin textures of the surface Dirac cones in the topological insulators}

To highlight the difference between the two representative Dirac semimetals and the topological insulator, here
we also provide a derivation of the orbital-resolved spin textures associated with the surface Dirac cones in the topological insulator.

To start, we directly consider the minimal continuum Hamiltonian describing the topological insulator, which reads
\begin{eqnarray}
\mathcal{H}_{\rm TI}(\bk)=[M(\bk_{s})+\frac{t_{z}}{2}k_{z}^{2}]\sigma_{z}s_{0}+\lambda(k_{x}\sigma_{x}s_{x}+k_{y}\sigma_{x}s_{y}+k_{z}\sigma_{z}s_{z}).
\end{eqnarray}
Also focusing on the bottom $z$-normal surface, we first determine the low-energy surface Hamiltonian.
Similarly, the first step is to decompose the Hamiltonian into two parts, i.e.,
$\mathcal{H}_{\rm TI}=\mathcal{H}_{0}+\mathcal{H}_{1}$, with
\begin{eqnarray}
\mathcal{H}_{0}(\bk)&=&[M(\bk_{s})+\frac{t_{z}}{2}k_{z}^{2}]\sigma_{z}s_{0}+\lambda k_{z}\sigma_{x}s_{z},\nonumber\\
\mathcal{H}_{1}(\bk)&=&\lambda(k_{x}\sigma_{x}s_{x}+k_{y}\sigma_{x}s_{y}).
\end{eqnarray}
Next replacing $k_{z}$ by $-i\partial_{z}$ and  solving the eigenvalue equation $\mathcal{H}_{0}(\bk_{s},-i\partial_{z})\varphi(x,y,z)=E(\bk_{s})\varphi(x,y,z)$ under
the boundary conditions $\varphi(x,y,z=0)=\varphi(x,y,z\rightarrow+\infty)=0$, one can find two zero-energy solutions of the form
\begin{eqnarray}
\varphi_{\alpha}^{b}(x,y,z)=\mathcal{N}\sin(\gamma_{1}z)e^{-\gamma_{2}z}e^{i(k_{x}x+k_{y}y)}\xi_{\alpha}^{b},\label{wf}
\end{eqnarray}
where $\alpha=\pm$, $\gamma_{1}=\sqrt{-2t_{z}M(\bk_{s})-\lambda^{2}}/t_{z}$, $\gamma_{2}=\lambda/t_{z}$,
$\mathcal{N}=2\sqrt{|\gamma_{2}(\gamma_{1}^{2}+\gamma_{2}^{2})/\gamma_{1}^{2}|}$ is the normalization constant, and $\xi_{\alpha}^{b}$ satisfy
$\sigma_{y}s_{z}\xi_{\alpha}^{b}=-\xi_{\alpha}^{b}$. The normalizability of the wave functions indicates that
the zero-energy surface states exist within the region
$M(\bk_{s})<0$. It is worth noting that  now the surface states exist at $\bk_{s}=\mathbf{0}$.

Choosing $\xi_{\pm}^{b}=|\sigma_{y}=\mp1,s_{z}=\pm1\rangle$ and considering
the basis to be $(\varphi_{+}^{b},\varphi_{-}^{b})^{T}$, the low-energy surface Hamiltonian
can be obtained by projecting $\mathcal{H}_{1}$ onto the basis, which is found
to take the form
\begin{eqnarray}
\mathcal{H}_{b}(\bk_{s})=\lambda\left(
                           \begin{array}{cc}
                             0 & k_{y}+ik_{x} \\
                             k_{y}-ik_{x} & 0 \\
                           \end{array}
                         \right)=\lambda(k_{y}\rho_{x}-k_{x}\rho_{y}).
\end{eqnarray}
The above low-energy surface Hamiltonian is well-known. Unlike the situation
for the opposite-parity Dirac semimetal, now the linear surface Dirac cone carry
the Dirac node  as the surface states exist
at $\bk_{s}=\mathbf{0}$.

Similarly, we again focus on the upper half cone and determine the orbital-resolved spin textures.
The eigenstate of the upper band of the low-energy surface Hamiltonian is of the form
$(1,-ie^{\phi_{\bk_{s}}})^{T}/\sqrt{2}$, accordingly, we have
\begin{eqnarray}
|w_{b}(\bk_{s})\rangle=\frac{1}{\sqrt{2}}(\xi_{+}^{b}-ie^{i\phi_{\bk_{s}}}\xi_{-}^{b}).
\end{eqnarray}
Based on the spinor $|w_{b}(\bk_{s})\rangle$, the orbital-resolved spin texture can be
determined by calculating
\begin{eqnarray}
\bar{s}_{i}^{b,o_{\pm}}(\bk_{s})=\langle w_{b}(\bk_{s})|\frac{(\sigma_{0}\pm\sigma_{z})}{2}s_{i}|w_{b}(\bk_{s})\rangle.
\end{eqnarray}
A straightforward calculation shows that
\begin{eqnarray}
\bar{s}_{x}^{b,0}(\bk_{s})&=&\langle w^{b}(\bk_{s})|\sigma_{0}s_{x}|w^{b}(\bk_{s})\rangle\nonumber\\
&=&\frac{1}{2}[(\xi_{+}^{b})^{\dag}\sigma_{0}s_{x}\xi_{+}^{b}+(\xi_{-}^{b})^{\dag}\sigma_{0}s_{x}\xi_{-}^{b}]
+\frac{1}{2}[ie^{-i\phi_{\bk_{s}}}(\xi_{-}^{b})^{\dag}\sigma_{0}s_{x}\xi_{+}^{b}-ie^{i\phi_{\bk_{s}}}
(\xi_{+}^{b})^{\dag}\sigma_{0}s_{x}\xi_{-}^{b}]\nonumber\\
&=&0,\nonumber\\
\bar{s}_{y}^{b,0}(\bk_{s})&=&\langle w^{b}(\bk_{s})|\sigma_{0}s_{y}|w^{b}(\bk_{s})\rangle
=0,\nonumber\\
\bar{s}_{z}^{b,0}(\bk_{s})&=&\langle w^{b}(\bk_{s})|\sigma_{0}s_{z}|w^{b}(\bk_{s})\rangle
=0,
\end{eqnarray}
and
\begin{eqnarray}
\bar{s}_{x}^{b,z}(\bk_{s})&=&\langle w^{b}(\bk_{s})|\sigma_{z}s_{x}|w^{b}(\bk_{s})\rangle\nonumber\\
&=&\frac{1}{2}[(\xi_{+}^{b})^{\dag}\sigma_{z}s_{x}\xi_{+}^{b}+(\xi_{-}^{b})^{\dag}\sigma_{z}s_{x}\xi_{-}^{b}]
+\frac{1}{2}[ie^{-i\phi_{\bk_{s}}}(\xi_{-}^{b})^{\dag}\sigma_{z}s_{x}\xi_{+}^{b}-ie^{i\phi_{\bk_{s}}}
(\xi_{+}^{b})^{\dag}\sigma_{z}s_{x}\xi_{-}^{b}]\nonumber\\
&=&\frac{i}{2}e^{-i\phi_{\bk_{s}}}-\frac{i}{2}e^{i\phi_{\bk_{s}}}\nonumber\\
&=&\sin\phi_{\bk_{s}},\nonumber\\
\bar{s}_{y}^{b,z}(\bk_{s})&=&\langle w^{b}(\bk_{s})|\sigma_{z}s_{y}|w^{b}(\bk_{s})\rangle\nonumber\\
&=&\frac{1}{2}[(\xi_{+}^{b})^{\dag}\sigma_{z}s_{y}\xi_{+}^{b}+(\xi_{-}^{b})^{\dag}\sigma_{z}s_{y}\xi_{-}^{b}]
+\frac{1}{2}[ie^{-i\phi_{\bk_{s}}}(\xi_{-}^{b})^{\dag}\sigma_{z}s_{y}\xi_{+}^{b}-ie^{i\phi_{\bk_{s}}}
(\xi_{+}^{b})^{\dag}\sigma_{z}s_{y}\xi_{-}^{b}]\nonumber\\
&=&-\frac{1}{2}(e^{-i\phi_{\bk_{s}}}+e^{i\phi_{\bk_{s}}})\nonumber\\
&=&-\cos\phi_{\bk_{s}},\nonumber\\
\bar{s}_{z}^{b,z}(\bk_{s})&=&\langle w^{b}(\bk_{s})|\sigma_{z}s_{z}|w^{b}(\bk_{s})\rangle\nonumber\\
&=&\frac{1}{2}[(\xi_{+}^{b})^{\dag}\sigma_{z}s_{z}\xi_{+}^{b}+(\xi_{-}^{b})^{\dag}\sigma_{z}s_{z}\xi_{-}^{b}]
+\frac{1}{2}[ie^{-i\phi_{\bk_{s}}}(\xi_{-}^{b})^{\dag}\sigma_{z}s_{z}\xi_{+}^{b}-ie^{i\phi_{\bk_{s}}}
(\xi_{+}^{b})^{\dag}\sigma_{z}s_{z}\xi_{-}^{b}]\nonumber\\
&=&0.
\end{eqnarray}
Accordingly, we have
\begin{eqnarray}
\bar{s}_{x}^{b,o_{+}}(\bk_{s})&=&\frac{1}{2}(\bar{s}_{x}^{b,0}(\bk_{s})+\bar{s}_{x}^{b,z}(\bk_{s}))
=\frac{1}{2}\sin\phi_{\bk_{s}},\nonumber\\
\bar{s}_{y}^{b,o_{+}}(\bk_{s})&=&\frac{1}{2}(\bar{s}_{y}^{b,0}(\bk_{s})+\bar{s}_{y}^{b,z}(\bk_{s}))
=-\frac{1}{2}\cos\phi_{\bk_{s}},\nonumber\\
\bar{s}_{z}^{b,o_{+}}(\bk_{s})&=&0,\nonumber\\
\end{eqnarray}
and
\begin{eqnarray}
\bar{s}_{x}^{b,o_{-}}(\bk_{s})&=&\frac{1}{2}(\bar{s}_{x}^{b,0}(\bk_{s})-\bar{s}_{x}^{b,z}(\bk_{s}))
=-\frac{1}{2}\sin\phi_{\bk_{s}},\nonumber\\
\bar{s}_{y}^{b,o_{-}}(\bk_{s})&=&\frac{1}{2}(\bar{s}_{y}^{b,0}(\bk_{s})-\bar{s}_{y}^{b,z}(\bk_{s}))
=\frac{1}{2}\cos\phi_{\bk_{s}},\nonumber\\
\bar{s}_{z}^{b,o_{-}}(\bk_{s})&=&0.
\end{eqnarray}
It is easy to see that the spin textures for the two orbitals are opposite, which is a natural result
due to the spin-orbit coupling. From the above expressions, one can see that the orbital-resolved
spin textures only depend one phase angle, i.e., $\phi_{\bk_{s}}$. Accordingly, the two orbital-resolved
spin textures carry the same winding number which is equal to one.

\end{widetext}

\end{document}